\documentclass[aps,prb,twocolumn,superscriptaddress,floatfix,notitlepage,a4paper,10pt]{revtex4-2}

\usepackage{xcolor}
\usepackage{graphicx}
\usepackage{pdfsync}
\usepackage{hyperref}
\hypersetup{colorlinks, linkcolor={red}, citecolor={blue}, urlcolor={blue}}
\usepackage{bm}
\usepackage{comment}
\usepackage[normalem]{ulem}

\begin{document}

\title{Microscopic theory of plasmon-enabled resonant terahertz detection in bilayer graphene}
\date{\today}

\begin{abstract}
The electron gas hosted in a two-dimensional solid-state matrix, such as a quantum well or a two-dimensional van der Waals heterostructure, supports the propagation of plasma waves.
Nonlinear interactions between plasma waves, due to charge conservation and current convection, generate a constant density gradient which can be detected as a dc potential signal at the boundaries of the system.
This phenomenon is at the heart of a plasma-wave photodetection scheme which was first introduced by Dyakonov and Shur for electronic systems with a parabolic dispersion and then extended to the massless Dirac fermions in graphene.
In this work, we develop the theory of plasma-wave photodetection in bilayer graphene, which has the peculiarity that the dispersion relation depends locally and dynamically on the intensity of the plasma wave.
In our analysis, we show how quantum capacitance effects, arising from the local fluctuations of the electronic dispersion, modify the intensity of the photodetection signal.
An external electrical bias, e.g.~induced by top and bottom gates, can be used to control the strength of the quantum capacitance corrections, and thus the photoresponse.
\end{abstract}

%\pacs{72.30.+q,73.21.-b,71.10.-w}

\author{Andrea Tomadin}
\affiliation{Dipartimento di Fisica dell'Università di Pisa, Largo Bruno Pontecorvo 3, I-56127 Pisa, Italy}

\author{Matteo Carrega}
\affiliation{CNR-SPIN,  Via  Dodecaneso  33,  16146  Genova, Italy}
    
\author{Marco Polini}
\affiliation{Dipartimento di Fisica dell'Università di Pisa, Largo Bruno Pontecorvo 3, I-56127 Pisa, Italy}
\affiliation{School of Physics \& Astronomy, University of Manchester, Oxford Road, Manchester M13 9PL, United Kingdom}
\affiliation{Istituto Italiano di Tecnologia, Graphene Labs, Via Morego 30, I-16163 Genova, Italy}

\maketitle

\section{Introduction}
\label{sect:introduction}

An electron system in a solid-state matrix displays collective density oscillations supported by the Coulomb repulsion between electrons.
If the period $T$ of such oscillations is much longer than the time $\tau_{\rm eq}$ that is needed for establishing a local thermal equilibrium in the electron system, then the oscillations are
described by a wave equation and are called \emph{plasma waves}.~\cite{Giuliani_and_Vignale}

As observed in a seminal paper~\cite{dyakonov_prl_1993} by Dyakonov and Shur (DS), a setup which allows to meet the condition $\tau_{\rm eq} \ll T$ is that of a field-effect transistor (FET), i.e.~a conductive channel where electrons roam, with a source and a drain contact, and capacitively coupled to a gate conductor.
The charges on the gate screen the density oscillations in the channel and reduce the intensity of the Coulomb interaction, lowering the frequency of the oscillations below the threshold $2\pi / \tau_{\rm eq}$.
In this regime, the electron system is well-described by hydrodynamic equations, which are nonlinear in the coupling between the electron density and velocity.
These hydrodynamic nonlinearities originate intriguing interference effects between the propagating plasma waves, as further pointed out by DS.~\cite{dyakonov_prb_1995,dyakonov_ieee_1996a,dyakonov_ieee_1996b}

Among these effects, one of notable practical importance is produced by subjecting the FET to a specific driving: feeding an ac potential between the source contact and the gate, with the drain contact floating.
This ac potential can be fed into the FET by an appropriately connected antenna.
The result is that a dc potential, i.e.~a \emph{photovoltage}, is established between source and drain, as a consequence of a standing wave being supported in the FET channel, which acts as a cavity.
This concept is called DS plasma-wave photodetection scheme, and has received steady theoretical and experimental interest for a few decades, especially in relation to the generation and detection of terahertz radiation.~\cite{knap_jimtw_2009, koppens_natnanotechnol_2014}

The theory predicts that resonant (i.e.~frequency-resolved) photodetection can be achieved, if it is possible to tune the plasma-wave speed, for example by electrical doping, and if the channel is sufficiently clean (i.e.~if the electronic momentum scattering rate is not much shorter than $T$).
Graphene, a two-dimensional crystal made of Carbon atoms,~\cite{KatsnelsonBook} possesses both these qualities: an electrically tunable carrier density,~\cite{novoselov_science_2004} and a large room-temperature mobility, especially if encapsulated in hexagonal boron nitride (hBN).~\cite{dean_natnanotech_2010} 
In particular, it has been shown that hBN-encapsulated graphene allows the electron system to sustain long-lived plasma excitations~\cite{woessner_natmater_2015,ni_nature_2018} and to enter the hydrodynamic regime~\cite{bandurin_science_2016, crossno_science_2016, ghahari_prl_2016, gallagher_science_2019, berdyugin_science_2019, polini_physicstoday_2020} even in the absence of a gate.
Indeed, graphene-based FETs have been identified early-on as ideal candidates to investigate the DS photodetection scheme~\cite{vicarelli_naturemater_2012, tredicucci_ieee_2014}.
Later, larger responsivity has been achieved~\cite{spirito_apl_2014} using bilayer-graphene (BLG)~\cite{KatsnelsonBook} channels and, finally, resonant DS photodetection has been demonstrated~\cite{bandurin_natcommun_2018} using hBN-encapsulated BLG.

\begin{figure}
\begin{center}
\includegraphics[width=3.46in]{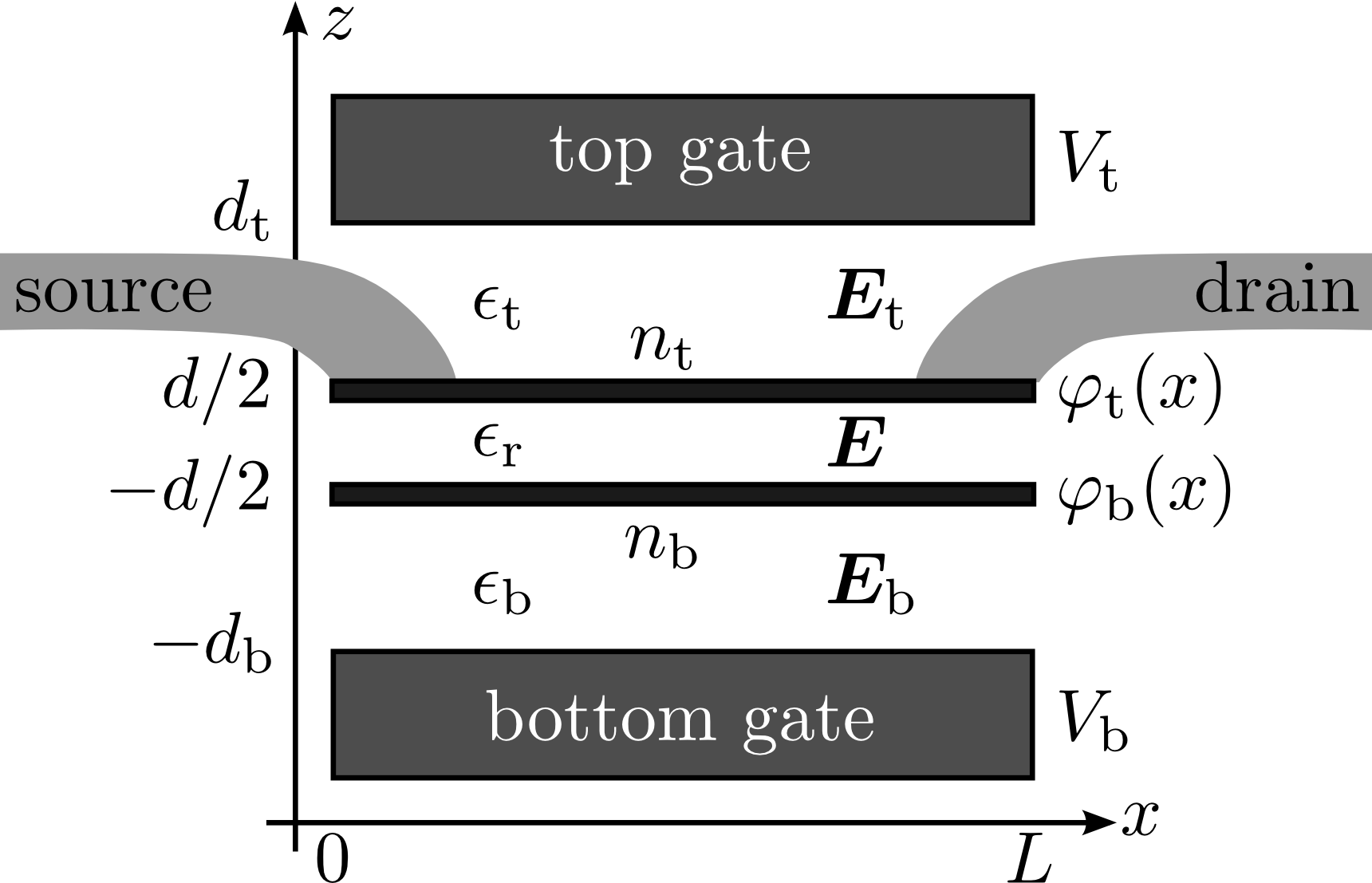}
\end{center}
\caption{\label{fig:setup}
Schematics of the setup.
The two graphene layers forming the bilayer system, at $z = -d/2$ and $z = d/2$, are displayed as thick black lines.
The top layer is contacted to source and drain, represented by the light shaded shapes.
The surfaces of the bottom and top gates, represented by dark shaded rectangles, lie at $z = -d_{\rm b}$ and $z = d_{\rm t}$.
The dielectric constants in the three regions of space delimited by the gates and the layers are denoted $\epsilon_{\rm b}$, $\epsilon_{\rm r}$, and $\epsilon_{\rm t}$ from bottom to top.
In the corresponding regions of space the electric field is denoted ${\bm E}_{\rm b}$, ${\bm E}$, and ${\bm E}_{\rm b}$.
The electric potential at the gates and along the layers is denoted $V_{\rm b}$, $V_{\rm t}$, $\varphi_{\rm b}(x)$, and $\varphi_{\rm t}(x)$.
Translational invariance is assumed in the $y$ direction. }
\end{figure}

Motivated by this recent experimental breakthrough, in this paper we formulate the theory of DS plasma-wave photodetection in dual-gated bilayer-graphene FETs.
We emphasize that, in Ref.~\cite{bandurin_natcommun_2018}, from a theory point of view, BLG was modeled as a Fabry-P\'erot resonator with effective parameters.
Such treatment does not yield a full picture of the dependence of the photovoltage on the experimental tuning knobs, such as the electron density and the gate voltages.
Here, we develop a theory that starts out from a microscopic description of electrons in BLG in terms of a two-bands Hamiltonian, and derive an expression for the DS photovoltage by means of electron hydrodynamic equations which are specific to BLG.
Indeed, in the electron hydrodynamic regime,~\cite{tomadin_prl_2014, torre_prb_2015, narozhny_prb_2015, levitov_natphys_2016, narozhny_annphys_2017, guo_pnas_2017, gabbana_prl_2018, shytov_prl_2018, ledwith_prl_2019}, the hydrodynamic equations in graphene feature peculiar terms~\cite{rudin_ijhses_2011, svintsov_jap_2012, tomadin_prb_2013} arising from its linear electronic dispersion.
In BLG, the linear electronic bands of two adjacent graphene sheets hybridize and give origin to a new electronic dispersion characterized by a position-dependent gap.~\cite{mccann_prb_2006,fogler_prb_2010,mccann_repprogphys_2013}
In this work, we show that the electron hydrodynamic equations in BLG are augmented by terms which derive from the finite gap in the electronic dispersion.
The electron density and gap fluctuations are coupled, which modifies the plasma-wave propagation and affects the dc photovoltage of a FET based on a bilayer-graphene channel.
We thus obtain, on a microscopic footing, the parametric dependence of the DC photovoltage on the dual-gated BLG FET.

We stress that BLG is different from a double- or multi-layer graphene system, where vertically-stacked graphene sheets are electronically decoupled, and the electronic dispersion in each sheet is unperturbed.
Double-layer setups support intra- and inter-layer plasmon modes~\cite{profumo_prb_2012, carrega_njp_2012, principi_prb_2012} and feature interesting interplay between plasma-wave propagation~\cite{ryzhii_jap_2013} and inter-layer electron tunneling or scattering.~\cite{ryzhii_apl_2013, ryzhii_jpd_2013, enaldiev_prb_2017, devega_acsphotonics_2017, guerrero_prb_2019}
Electron tunneling in multi-layer graphene~\cite{ryzhii_jap_2014, ryzhii_2dmaterials_2015} or in gate-defined lateral tunnel junctions in BLG~\cite{gayduchenko_arxiv_2020} can also be exploited to achieve photodetection mechanisms different from the DS scheme.

The paper is organized as follows.
In Sec.~\ref{sect:setup} we introduce the model and define the set of variables describing the electron system.
For the sake of clarity, we introduce all degrees of freedom (quantum, thermodynamic, electrostatic) in full detail, and the relations between them in focused sub-sections, aiming at a self-contained and pedagogical exposition.
In Sec.~\ref{sec:hydrodynamicequations} we introduce the dynamical model coupling quantum, thermodynamic, and electrostatic variables, which takes the form of a set of hydrodynamic equations.
In Sec.~\ref{sec:plasmawaves} we discuss the linearized solutions of the hydrodynamic equations, which describe plasma waves, both freely propagating and in the presence of boundary conditions.
In Sec.~\ref{sect:photovoltage} we calculate the photovoltage when the electron system in a FET channel is subjected to the Dyakonov-Shur boundary conditions.
The differences of the photoresponse function with respect to the single-layer graphene case are elucidated.
The broadband limit of the photoresponse function is discussed in the Appendix.
Finally, in Sec.~\ref{sec:summary} we summarize our procedure and findings.

\section{Electron density in dual-gated bilayer graphene}

\subsection{The setup}
\label{sect:setup}

We model BLG as two graphene sheets lying in the planes $z = -d / 2$ (bottom layer) and $z = d / 2$ (top layer), extended along the $x$ direction from $x = 0$ to $x = L$.
(See Fig.~\ref{fig:setup}.)
As anticipated in Sec.~\ref{sect:introduction}, and discussed in detail in Sec.~\ref{sect:bilayer}, the electronic states in the two sheets are hybridized.
We assume that the system is translationally invariant in the $y$ direction. 
Two perfect conductors lie in the planes $z = -d_{\rm b}$ (bottom gate) and $z = d_{\rm t}$ (top gate).
The uniform electric potential of the bottom (top) gate is $V_{\rm b}$ ($V_{\rm t}$).
The electric potential in the bottom (top) layer is $\varphi_{\rm b}(x)$ [$\varphi_{\rm t}(x)$].
The dielectric constant is: $\epsilon_{\rm b}$ for $-d_{\rm b}<z<-d/2$, $\epsilon_{\rm r}$ for $-d/2<z<d/2$, $\epsilon_{\rm t}$ for $d / 2 < z < d_{\rm t}$.
The electric field is: ${\bm E}_{\rm b}(x,z)$ for $-d_{\rm b}<z<-d/2$, ${\bm E}(x,z)$ for $-d/2<z<d/2$, ${\bm E}_{\rm t}(x,z)$ for $d / 2 < z < d_{\rm t}$.
At $x = 0$ ($x = L$) the top graphene layers touches the source (drain) contact.
We assume that the electric potential on the top graphene layer is locally equal to that on the contacts, while the potential on the bottom graphene layer is free to float and is determined as detailed in the following section.
Vectors in the two-dimensional (2D) $x-y$ space are denoted as ${\bm v} = (v_{x}, v_{y})$.

\subsection{Bilayer graphene Hamiltonian}
\label{sect:bilayer}

Electrons roaming in the BLG are described by the following Hamiltonian in the two-bands approximation~\cite{mccann_prb_2006,fogler_prb_2010,mccann_repprogphys_2013}
\begin{equation}\label{eq:blghamil}
\hat{\cal H}(x) = \left(
\begin{array}{cc}
0 & - \frac{(\hat{p}_{x} - i \hat{p}_{y})^2}{2m} \\
- \frac{(\hat{p}_{x} + i \hat{p}_{y})^2}{2m} & 0
\end{array}
\right) - \frac{\Delta(x)}{2}\left(
\begin{array}{cc}
1 &0 \\
0 &-1
\end{array}
\right)~.
\end{equation}
Here, $\hat{p}_{x}$ and $\hat{p}_{y}$ are the components of the electron momentum operator.
The matrices act in the layer space, where the first (second) component corresponds to the bottom (top) layer.
The index in the layer space is $\ell = {\rm b}$ (bottom layer) and $\ell = {\rm t}$ (top layer).
The kinetic part of the Hamiltonian uses the effective mass $m \equiv \gamma_{1} / (2 v_{\rm F}^{2})$, where $v_{\rm F}$ is the Fermi velocity of electrons in single-layer graphene and $\gamma_{1}$ is a hopping parameter of the tight-binding model~\cite{mccann_prb_2006,fogler_prb_2010,mccann_repprogphys_2013}.
The value of the effective mass is $m \simeq 0.035~m_{\rm e}$, where $m_{\rm e}$ is the bare electron mass.
Finally, the potential part of the Hamiltonian, which is diagonal in the layer space, represents the potential energy difference between the two graphene layers, also called \emph{asymmetry},
\begin{equation}\label{eq:deltadefinition}
\Delta(x) \equiv -e[\varphi_{\rm t}(x) - \varphi_{\rm b}(x)]~.
\end{equation}
A different potential energy on the two layers, resulting in $\Delta \neq 0$, can be induced by the electric field of a gate, for example.
The Hamiltonian~(\ref{eq:blghamil}) depends parametrically on the coordinate $x$ through $\Delta(x)$.
This approach is justified if the wave vector $k$ of the typical variation of $\Delta(x)$ is much smaller than the Fermi wave vector $k_{\rm F}$, i.e.~we require
\begin{equation}
k \ll k_{\rm F}~.
\end{equation}
The eigenstates of the Hamiltonian~(\ref{eq:blghamil}) are labelled by the wave vector ${\bm k}$ and the band index $\lambda$, which assumes the values $\lambda = +1$ (conduction band) and $\lambda = -1$ (valence band).
The eigenvalues are
\begin{equation}\label{eq:blgdispersion}
\varepsilon_{\lambda,{\bm k}}(x) = \lambda \varepsilon_{{\bm k}}(x), \quad
\varepsilon_{{\bm k}}(x) = \sqrt{\left \lbrack \frac{\hbar^{2} |{\bm k}|^2}{2m}\right \rbrack^2 + \frac{\Delta(x)^2}{4}}~,
\end{equation}
and the corresponding eigenvectors are
\begin{eqnarray}\label{eq:blgeigenvectors}
\label{eq:wf_2}
\psi_{\lambda,{\bm k}}(x) & = & \sqrt{\frac{\varepsilon_{\lambda,{\bm k}}(x) - \Delta(x)/2}{2\varepsilon_{\lambda,{\bm k}}(x)}}
\left ( \begin{array}{c}
1 \\ 
- \frac{e^{2 i \varphi_{\bm k}} (\hbar {\bm k})^2/(2m)}{\varepsilon_{\lambda, {\bm k}}(x) - \Delta(x)/2}
\end{array} \right )~, \nonumber \\
\end{eqnarray}
where $\varphi_{\bm k}$ is the angle between the vector ${\bm k}$ and the versor $\hat{\bm x}$, i.e.~$k_{x} + i k_{y} = |{\bm k}| e^{i \varphi_{\bm k}}$.
(Notice that the argument in the square root is always positive.)
From the dispersion~(\ref{eq:blgdispersion}) we see that the band gap corresponds to the parameter $|\Delta(x)|$.

\subsection{Local quasi-equilibrium probability distribution}

The electronic hydrodynamic regime~\cite{tomadin_prl_2014, torre_prb_2015, narozhny_prb_2015, levitov_natphys_2016, narozhny_annphys_2017, guo_pnas_2017, gabbana_prl_2018, shytov_prl_2018, ledwith_prl_2019} is formally captured by the assumption that the system is in a state of local quasi-equilibrium, i.e.~the probability that the single-particle state with wave vector ${\bm k}$, in the band $\lambda$, is occupied is given by a ``displaced'' Fermi-Dirac distribution $f_{\lambda,{\bm k}}$.~\cite{rudin_ijhses_2011,svintsov_jap_2012,tomadin_prb_2013}
For simplicity, we assume that the system is everywhere homopolar, i.e.~there is a single chemical potential for both conduction and valence band.
In this case, the probability distribution is
\begin{equation}\label{eq:distrohydro}
f_{\lambda,{\bm k}}(x) = 1 / \left \lbrace e^{\beta \lbrack \varepsilon_{\lambda,{\bm k}} - \hbar {\bm v}(x) \cdot {\bm k} - \mu(x) \rbrack} + 1 \right \rbrace~.
\end{equation}
The inverse temperature $\beta = 1 / (k_{\rm B} T)$ is assumed to be constant and homogeneous in the system.
We suppose that the chemical potential is everywhere much larger than the temperature, $\mu(x) \gg k_{\rm B} T$, which allows us to neglect the effects of local heating and horizontal transport due to temperature gradients. 
The vector ${\bm v}(x)$ is the local electron drift velocity and $\mu(x)$ is the local chemical potential.
Consistently with the assumption of translational invariance along $y$, in the following we assume the drift velocity to be directed along $x$ only.
We point out that the distribution~(\ref{eq:distrohydro}) is defined for electron states in the bilayer and does not discriminate between the bottom and the top layer, since the electron wave functions in the two layers are hybridized.

\subsection{Relation between the electron density and the Hamiltonian variables}

The 2D electron density in the $\ell$th layer is given by
\begin{equation}
\label{eq:density_ell}
n_{\ell}(x) = N_{\rm f} \int \frac{d^{2}{\bm k}}{(2\pi)^2} \sum_{\lambda = \pm 1} |\lbrack \psi_{\lambda,{\bm k}}(x) \rbrack_{\ell} |^2 f_{\lambda,{\bm k}}(x)~,
\end{equation}
where the factor $N_{\rm f} = 4$ corresponds to the spin and valley degeneracy in each layer and $[\dots]_{\ell}$ denotes the $\ell$th component of the eigenvector in Eq.~(\ref{eq:blgeigenvectors}).
The electron density polarization between the two layers is defined as
\begin{equation}
\label{eq:n1_minus_n2}
n_{\rm b}(x) - n_{\rm t}(x) =  N_{\rm f} \frac{\Delta(x)}{2} \int \frac{d^2{\bm k}}{(2\pi)^2}  \frac{1}{\varepsilon_{{\bm k}}(x)} \sum_{\lambda=\pm 1}f_\lambda ({\bm k},x)~.
\end{equation}
For $\Delta(x) = 0$ the top- and bottom-layer densities coincide, $n_{\rm t}(x) = n_{\rm b}(x)$.
The total electron density on the BLG is defined as the sum of the electron densities on the two layers
\begin{equation}\label{eq:totaldensity}
n(x) \equiv n_{\rm b}(x) + n_{\rm t}(x)~.
\end{equation}

The integrals in Eqs.~(\ref{eq:density_ell}) and ~(\ref{eq:n1_minus_n2}) can be calculated analytically in the limit of low temperature and small drift velocity, defined by the inequalities
\begin{equation}
v(x) \ll v_{\rm F}, \quad
k_{\rm B} T \ll \varepsilon_{\rm F}~,
\end{equation}
where $\varepsilon_{\rm F}$ is the Fermi energy and $v_{\rm F}$ is the Fermi velocity.
In this limit, the probability distribution~(\ref{eq:distrohydro}) simplifies to
\begin{equation}\label{eq:fermistep}
f_{\lambda,{\bm k}}(x) = \Theta \left \lbrack \varepsilon_{\rm F}(x) - \varepsilon_{\lambda, {\bm k}}(x)  \right \rbrack~,
\end{equation}
where $\Theta(x)$ is the unit step function and we have dropped the dependence on the drift velocity ${\bm v}(x)$ in the argument.
The contribution to the integral~(\ref{eq:n1_minus_n2}) due to the velocity is of second order in $|{\bm v}(x)| / v_{\rm F}$ and can be neglected in the derivation of the hydrodynamic equations (see Sec.~\ref{sec:hydrodynamicequations}).
To evaluate the contribution of the valence band, it is necessary to introduce a momentum cutoff.~\cite{mccann_repprogphys_2013}
The electron density polarization and the total density read, respectively:
\begin{eqnarray} \label{eq:densitydiff}
& & n_{\rm b}(x) - n_{\rm t}(x) = - \frac{n_{\perp}}{2 \gamma_{1}} \Delta(x) \times  \nonumber \\
& & \ln \left ( \frac{|n(x)|}{2 n_{\perp}} + \frac{1}{2} \sqrt{ \left \lbrack \frac{n(x)}{n_{\perp}} \right \rbrack^{2} + \left \lbrack \frac{\Delta(x)}{2 \gamma_{1}} \right \rbrack^{2} } \right )~,
\end{eqnarray}
\begin{equation} \label{eq:densitytot}
n(x) = \frac{2 m }{\hbar^{2} \pi}\sqrt{\varepsilon_{{\rm F}}(x)^2 - \frac{\Delta(x)^2}{4}}~,
\end{equation}
where $n_{\perp} \equiv 4 v_{\rm F}^{2} m^{2} / (\hbar^{2} \pi)$.

\subsection{Relation between the electron density and the electric potential}

To calculate the relation between the electron density and the electric potential, we consider first a homogeneous system, translationally invariant in the $x$ direction.
Therefore, for the sake of simplicity, in this section we drop the $x$ dependence from the variables.
Due to the translational invariance, the electric fields are uniform and directed along the $\hat{\bm z}$ direction:
\begin{equation}
{\bm E}_{\rm b}(z) = \hat{\bm z} E_{\rm b}, \quad
{\bm E}(z) = \hat{\bm z} E, \quad
{\bm E}_{\rm t}(z) = \hat{\bm z} E_{\rm t}~.
\end{equation}
From the straightforward application of Gauss's law to regions enclosed by planes orthogonal to the $z$ axis, we find the relations between the electric fields (in SI units) and the electron density:
\begin{eqnarray}
\label{eq:gauss1}
- \epsilon_0 \epsilon_r E  + \epsilon_0 \epsilon_t E_t & = & - e n_{\rm t}, \nonumber \\
- \epsilon_0 \epsilon_b E_b  + \epsilon_0 \epsilon_r E & = & - e n_{\rm b} ~.
\end{eqnarray}

It is convenient to find the relations between the uniform electric fields and the electric potentials on the gates and the graphene layers.
The magnitude $E$ of the electric field between the two layers is
\begin{equation} \label{eq:deltaelectric}
E = -(\varphi_{\rm t} - \varphi_{\rm b}) / d = \Delta / (ed)~,
\end{equation}
where the second equality follows from Eq.~(\ref{eq:deltadefinition}).
The magnitude $E_{\rm t}$ ($E_{\rm b}$) of the top (bottom) electric field can be determined using the electric potential $V_{\rm t}$ ($V_{\rm b}$) on the top (bottom) gate.
We find that
\begin{equation}
E_{\rm t} = -(V_{\rm t} - \varphi_{\rm t}) / (d_{\rm t} - d/2), \quad
E_{\rm b} = -(\varphi_{\rm b} - V_{\rm b}) / (d_{\rm b} - d/2)~.
\end{equation}
It is now convenient to define the \emph{gate to channel swing} $U_{{\rm t}}$, $U_{{\rm b}}$ for the top and bottom layer, respectively, and the sum of the two:
\begin{equation}\label{eq:swings}
U_{\rm t} \equiv V_{\rm t} - \varphi_{\rm t}, \quad
U_{\rm b} \equiv V_{\rm b} - \varphi_{\rm b}, \quad
U \equiv U_{\rm t} + U_{\rm b}~.
\end{equation}
From the above definitions we find:
\begin{eqnarray}\label{eq:electricfromswing}
E_{\rm t} & = & - U_{\rm t} / (d_{\rm t} - d/2), \quad
E_{\rm b} =  U_{\rm b} / (d_{\rm b} - d/2)~, \\
\Delta & = & -e [ V_{\rm t} - V_{\rm b} - U_{\rm t} + U_{\rm b} ]~.
\end{eqnarray}
We note that, in principle, one could use the latter equation to eliminate $\Delta$ using $U_{\rm t}$ and $U_{\rm b}$.
However, as we see below, it is convenient to use this equation to solve for $U_{\rm b}$ and find $\Delta$ from the implicit relation~(\ref{eq:deltafromdensity}).
We also note that the results depend only on the difference between the gate potentials, as expected.

To rewrite the above equations in a more transparent form, it is convenient to define the capacitance per unit area of the three capacitors formed by the two layers, and the top (bottom) layer with the top (bottom) gate:
\begin{eqnarray}
C_{\rm r} & = & \epsilon_0\epsilon_{{\rm r}} / d, \nonumber \\
C_{{\rm t}} & = & \epsilon_0\epsilon_{{\rm t}} / (d_{{\rm t}} - d/2), \nonumber \\
C_{{\rm b}} & = & \epsilon_0\epsilon_{{\rm b}} / (d_{{\rm b}} - d/2)~.
\end{eqnarray}
Then, from Eqs.~(\ref{eq:gauss1}) and (\ref{eq:electricfromswing}) we finally find the electron density as a function of the gate-to-channel swings and the Hamiltonian parameter $\Delta$
\begin{equation}
n_{\rm t} = \frac{C_{\rm t}}{e} U_{\rm t} + \frac{C_{\rm r}}{e}\frac{\Delta}{e},\quad
n_{\rm b} = \frac{C_{\rm b}}{e} U_{\rm b} - \frac{C_{\rm r}}{e}\frac{\Delta}{e}~.
\end{equation}
Taking the difference and the sum of the latter expression we also find the electron density polarization and the total electron density, respectively:
\begin{equation} \label{eq:polarpoten}
n_{\rm b} - n_{\rm t} = - \frac{C_{\rm t}}{e} U_{\rm t} + \frac{C_{\rm b}}{e} U_{\rm b} - 2\frac{C_{\rm r}}{e}\frac{\Delta}{e}~,
\end{equation}
\begin{equation} \label{eq:denstotpoten}
n = \frac{C_{\rm t}}{e} U_{\rm t} + \frac{C_{\rm b}}{e} U_{\rm b}~.
\end{equation}

\subsection{Local capacitance approximation}
\label{ssec:gca}

We consider now the case of a non-uniform system where the electric potentials and the electron density varies on a length scale $l$ which is much larger than the vertical dimension of the system, i.e.~$l \gg d_{\rm t} + d_{\rm b}$.
In this case, we assume that the relations (\ref{eq:polarpoten}) and (\ref{eq:denstotpoten}) between the densities and the gate-to-channel swings hold \emph{locally} in space for any $x$.
This approximation is usually referred to as the ``local capacitance'' or ``gradual channel'' approximation~\cite{dyakonov_prl_1993} and reads:
\begin{equation}\label{eq:densitydiffswings}
n_{\rm b}(x) - n_{\rm t}(x) = \frac{C_{\rm b}}{e} U_{\rm b}(x) - \frac{C_{\rm t}}{e} U_{\rm t}(x) - 2\frac{C_{\rm r}}{e}\frac{\Delta(x)}{e}~,
\end{equation}
\begin{equation}\label{eq:densitytotswings}
n(x) = \frac{C_{\rm t}}{e} U_{\rm t}(x) + \frac{C_{\rm b}}{e} U_{\rm b}(x)~,
\end{equation}
We also consider the same approximation for  $\Delta(x)$, obtaining
\begin{equation}\label{eq:deltaswings}
\Delta(x) = -e [ V_{\rm t} - V_{\rm b} - U_{\rm t}(x) + U_{\rm b}(x) ]~.
\end{equation}

\subsection{Determination of the variables as a function of the total density}
\label{sect:systemsolution}

The equations ${\cal E} = \{$(\ref{eq:densitydiff}), (\ref{eq:densitytot}), (\ref{eq:densitydiffswings}), (\ref{eq:densitytotswings}), and (\ref{eq:deltaswings})$\}$ are a system of five independent equations for the six variables: ${\cal V}$ = $\{ n(x)$, $n_{\rm b}(x)-n_{\rm t}(x)$, $\Delta(x)$, $\varepsilon_{\rm F}(x)$, $U_{\rm t}(x)$, and $U_{\rm b}(x) \}$.
This means that any variable in a subset of five variables of ${\cal V}$ can be expressed as a function of the remaining sixth variable in ${\cal V}$.
In other words, a single variable is sufficient to uniquely determine the state of the system at each point $x$ in space.
We choose such variable to be the total density $n(x)$.
We present now a convenient sequence of substitutions to determine all the other variables in terms of $n(x)$.
We can directly solve (\ref{eq:densitytot}) for $\varepsilon_{\rm F}(x)$ and (\ref{eq:deltaswings}) for $U_{\rm b}(x)$:
\begin{equation}
\varepsilon_{\rm F}(x) = \sqrt{ \left \lbrack \frac{\hbar^{2} \pi n(x)}{2 m} \right \rbrack^{2} + \frac{\Delta(x)^{2}}{4}}~,
\end{equation}
\begin{equation}\label{eq:swingb}
U_{\rm b}(x) = U_{\rm t}(x) - \frac{\Delta(x)}{e} - (V_{\rm t} - V_{\rm b})~.
\end{equation}
We point out that the choice to eliminate $U_{\rm b}(x)$ instead of $U_{\rm t}(x)$ arises from the fact that the boundary conditions (BCs) for the hydrodynamic equations of motion (see Sec.~\ref{sec:hydrodynamicequations}) will be imposed on the top layer, i.e.~the one which touches the contacts.
Then, from (\ref{eq:densitytotswings}) we obtain $U_{\rm t}(x)$ in terms of $n(x)$ and $\Delta(x)$
\begin{equation}\label{eq:swingtdens}
\frac{C_{\rm t} + C_{\rm b}}{e} U_{\rm t}(x) =  n(x) + \frac{C_{\rm b}}{e} \frac{\Delta(x)}{e} + \frac{C_{\rm b}}{e}(V_{\rm t} - V_{\rm b})~.
\end{equation}
Finally, we eliminate $n_{\rm b}(x) - n_{\rm t}(x)$ by equating the right-hand side of Eqs.~(\ref{eq:densitydiff}) and (\ref{eq:densitydiffswings}).
In the equation that we find in this way, we substitute the expressions for $\varepsilon_{\rm F}(x)$, $U_{\rm b}(x)$, and $U_{\rm t}(x)$.
The result is an algebraic equation with contains $\Delta(x)$ and $n(x)$ only
\begin{eqnarray}\label{eq:deltafromdensity}
0 & = & n(x) \frac{C_{\rm t} - C_{\rm b}}{C_{\rm t} + C_{\rm b}}
+ \frac{V_{\rm t} - V_{\rm b}}{e} \frac{2 C_{\rm b} C_{\rm t}}{C_{\rm t} + C_{\rm b}} - \nonumber \nonumber \\
& & 2 \frac{\Delta(x)}{e^{2}} \frac{C_{\rm b} C_{\rm r} + C_{\rm b} C_{\rm t} + C_{\rm r} C_{\rm t}}{C_{\rm t} + C_{\rm b}} + \nonumber \\
& & \frac{n_{\perp}}{2 \gamma_{1}} \Delta(x) \ln{\left ( \frac{|n(x)|}{2 n_{\perp}} + \frac{1}{2} \sqrt{ \left \lbrack \frac{n(x)}{n_{\perp}} \right \rbrack^{2} + \left \lbrack \frac{\Delta(x)}{2 \gamma_{1}} \right \rbrack^{2} } \right )}~. \nonumber \\
\end{eqnarray}
Eq.~(\ref{eq:deltafromdensity}) is the main result of the lengthy derivation presented in this section.
It is the \emph{effective constitutive equation} of our model, which connects the microscopic Hamiltonian parameter $\Delta(x)$ to the macroscopic variable $n(x)$.

To understand the physical content of Eq.~(\ref{eq:deltafromdensity}), let us consider the limit $\Delta(x) \gg \hbar^{2} \pi n(x) / m$.
Using $\log ( \sqrt{\xi^{2} + 1} + \xi ) \sim \xi$ for $\xi \ll 1$, with $\xi = \hbar^{2} \pi n(x) / [m \Delta(x)]$, we see that the last term in the equation is $\sim n(x)$.
Then, neglecting $n(x)$ in the first term of the equation with respect to $\Delta (x)$, we find a constant value for $\Delta$ given by
\begin{eqnarray}
C_{\rm r} \Delta(x) & = & - e ( V_{\rm t} - V_{\rm b} ) C_{\rm series}, \nonumber \\
C_{\rm series} & = & \frac{1}{C_{\rm r}^{-1} + C_{\rm t}^{-1} + C_{\rm b}^{-1}}~,
\end{eqnarray}
where $C_{\rm series}$ is the series capacitance of the three capacitors formed by the gates and the two layers.
In this case, Eq.~(\ref{eq:deltafromdensity}) reduces to the requirement that the total charge on the gates must neutralize the total charge on the BLG.
In the general case, Eq.~(\ref{eq:deltafromdensity}) describes how the two gates, by screening the carrier density $n(x)$, induce an energy potential difference $\Delta(x)$ between the two graphene layers.
This mechanism has first been discussed by McCann et al. in Ref.~\cite{mccann_prb_2006}.

We note that the system ${\cal E}$, although nonlinear, features the following scaling relation: if the electric potential difference $V_{\rm t} - V_{\rm b}$ is multipled by a dimensionless factor, then the system is solved by multiplying all variables by the same factor as well.
This means that the electric potential difference between the top and the bottom gate merely sets the scale of the fields in the bilayer system, and can be fixed to a reasonable reference value in the analysis.
In the following we use $V_{\rm t} - V_{\rm b} = 100~{\rm V}$ for definiteness.

Solving the system ${\cal E}$, one obtains the functional dependence of all the other variables in terms of $n$.
The density-dependence of $U_{\rm t}(n)$ and $\Delta(n)$ is exemplified in Figs.~\ref{fig:electrostaticsthick} and~\ref{fig:electrostaticsgate} in a range of parameters.
Since the density $n(x)$ depends on the position $x$, the solution also yields the spatial dependence of the variables, which we denote e.g.~$U_{\rm t}(x) = U_{\rm t}(n(x))$.
Similarly, the notation $n(x) = n(U_{\rm t}(x))$ means that one has to solve the system ${\cal E}$ given the value of $U_{\rm t}(x)$ and calculate the remaining variables.
More generally, any variable $\chi$ in ${\cal V}$ can be used as independent variable and the remaining five variables in ${\cal V}$ can be expressed as functions of $\chi$.
We will use this notation in the following when convenient.

\subsection{Linearization around a homogeneous state}
\label{sect:elstaticslinear}

The system ${\cal E}$ can be solved at each point in space $x$ and at each instant in time $t$, to obtain the instantaneous values of the variables ${\cal V}$ in the whole system.
However, if the system is almost homogeneous, i.e.~$n(x) \simeq \bar{n}$, the spatial profile of all variables can be obtained by calculating the spatial fluctuations of one variable $\chi$ around its equilibrium value $\bar{\chi}$ and the derivative of the other variables with respect to $\chi$ at equilibrium.
For example, using the density as the independent variable:
\begin{equation}
U_{\rm t}(x) = U_{\rm t}(n(x)) \simeq U_{\rm t}(\bar{n}) + \left ( \frac{d U_{\rm t}}{d n} \right )_{n = \bar{n}} [ n(x) - \bar{n}] + \dots~. 
\end{equation}
Space- and time-derivative of the variables can also be expanded similarly, for example:
\begin{eqnarray}
& & \partial_{x} U_{\rm t}(x) = \partial_{x} U_{\rm t}(n(x)) =
\left ( \frac{d U_{\rm t}}{d n} \right )_{n = n(x)} \times \frac{\partial n(x)}{\partial x} \nonumber \\
& &  \simeq
\left \lbrace
\left ( \frac{d U_{\rm t}}{d n} \right )_{n = \bar{n}}
+ \left ( \frac{d^{2} U_{\rm t}}{d n^{2}} \right )_{n = \bar{n}} [ n(x) - \bar{n}]
+ \dots \right \rbrace \times
\frac{\partial n(x)}{\partial x}~. \nonumber \\
\end{eqnarray}

\begin{figure}
(a)\includegraphics[width=0.75\linewidth]{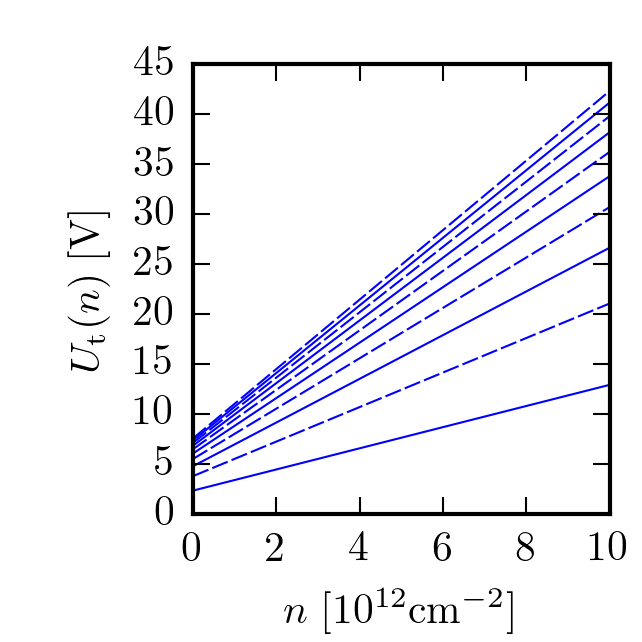}
(b)\includegraphics[width=0.75\linewidth]{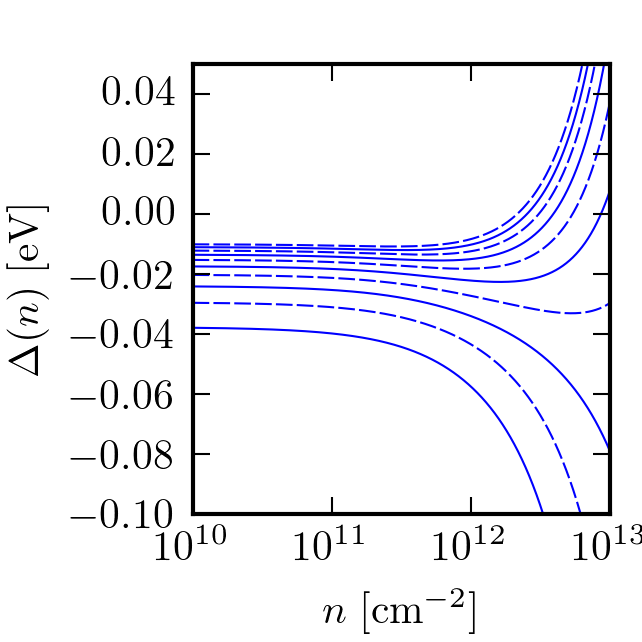}
\caption{\label{fig:electrostaticsthick}
(a) The gate to channel swing for the top layer and (b) the asymmetry, as a function of the density $n$.
Different lines correspond to values of $d_{\rm t}$ equally spaced from $d_{\rm t} = 30~{\rm nm}$ (solid) to $d_{\rm t} = 300~{\rm nm}$ (dashed).
The other parameters are: $d_{\rm b} = 100~{\rm nm}$, $\tau = 1~{\rm ps}$, $L = 5~\mu{\rm m}$, $V_{\rm T} - V_{\rm B} = 10~{\rm V}$. }
\end{figure}

\begin{figure}
(a)\includegraphics[width=0.75\linewidth]{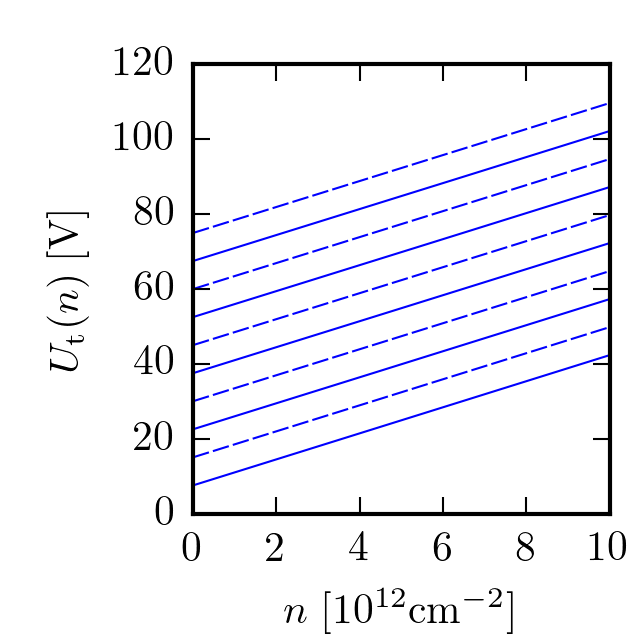}
(b)\includegraphics[width=0.75\linewidth]{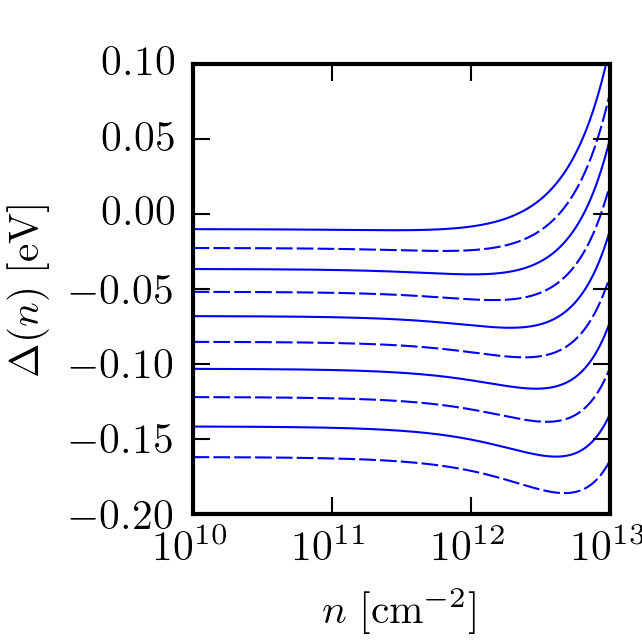}
\caption{\label{fig:electrostaticsgate}
Same as in Fig.~\ref{fig:electrostaticsthick}, but here
different lines correspond to values of $V_{\rm t} - V_{\rm b}$ equally spaced from $V_{\rm t} - V_{\rm b} = 10~{\rm V}$ (solid) to $V_{\rm t} - V_{\rm b} = 100~{\rm V}$ (dashed).
The other parameters are: $d_{\rm t} = 300~{\rm nm}$, $d_{\rm b} = 100~{\rm nm}$, $\tau = 1~{\rm ps}$, $L = 5~\mu{\rm m}$. }
\end{figure}

It is important to realize that, due to the nonlinearity of the system ${\cal E}$, the harmonic oscillation of any ``drive'' variable $\chi_{\rm d}$ at frequency $\omega$ produces anharmonic oscillations of the other variables, i.e.~a different variable $\chi$ in ${\cal V}$ has components oscillating at frequencies which are multiples of $\omega$, including the dc component with $\omega = 0$.
A component of $\chi$ oscillating with a frequency $\sum_{i = 1}^{n} \lambda_{i} \omega$, where $\lambda_{i} \in \{-1,1\}$, is obtained by expanding $\chi$ up to order $n$ in the drive $\chi_{\rm d}$.
Let us denote the expansion of $\chi$ as follows:
\begin{equation}\label{eq:expansion2o}
\chi(x,t) = \bar{\chi} + \chi^{(1)}(x,t) + \delta \chi(x) + \chi^{(2)}(x,t) + \dots~.
\end{equation}
Here, $\bar{\chi}$ represents the steady uniform value.
The components $\chi^{(1)}(x,t)$ and $\chi^{(2)}(x,t)$ oscillate at frequency $\omega$ and $2\omega$ and are first- and second-order in the drive, respectively.
Finally, $\delta \chi(x)$ is the second-order dc component.

For future purposes, let us specify the relations between some components of the density $n(x,t)$ and of the gate to channel swing for the top gate $U_{\rm t}(x,t)$, up to second order.
From now on, we use the prime symbol to denote derivative with respect to the density $n$ and the bar symbol to denote quantities evaluated at $n = \bar{n}$:
\begin{equation}\label{eq:notationderiv}
\bar{\chi} \equiv \chi(\bar{n}), \quad
\bar{\chi}' \equiv \left ( \frac{d \chi}{d n} \right )_{n = \bar{n}}~.
\end{equation}
Expanding the left- and right-hand side of Eq.~(\ref{eq:swingtdens}), and equating terms of the same order, we find
\begin{equation}\label{eq:expDensSwing0}
\bar{U}_{\rm t} = \frac{e}{C_{\rm t} + C_{\rm b}} \left \lbrack \bar{n} + \frac{C_{\rm b}}{e} \frac{\bar{\Delta}}{e} + \frac{C_{\rm b}}{e}(V_{\rm t} - V_{\rm b}) \right \rbrack~,
\end{equation}
for the homogeneous components,
\begin{eqnarray}\label{eq:expDensSwing1}
U_{\rm t}^{(1)}(x,t) & = & \bar{U}_{\rm t}' n^{(1)}(x,t), \nonumber \\
\quad \bar{U}_{\rm t}' & = & \frac{e}{C_{\rm t} + C_{\rm b}} \left ( 1 + \frac{C_{\rm b}}{e} \frac{\bar{\Delta}'}{e} \right )~,
\end{eqnarray}
for the first-order components, and
\begin{eqnarray}\label{eq:expDensSwing2}
\delta U_{\rm t}(x) & = & \bar{U}_{\rm t}' \delta n(x) + \frac{1}{2} \bar{U}_{\rm t}'' \langle n^{(1)}(x,t)^{2} \rangle_{t}, \nonumber \\
\quad \bar{U}_{\rm t}'' & = & \frac{C_{\rm b}}{C_{\rm t} + C_{\rm b}} \frac{\bar{\Delta}''}{e}~,
\end{eqnarray}
for the dc second-order components, where we have introduced the time-average over one period of the drive:
\begin{equation}\label{eq:timeaverage}
\langle f(t) \rangle_{t} = \frac{\omega}{2\pi}\int_0^{2\pi/\omega} d t f(t)~.
\end{equation}
The derivatives of $U_{\rm t}$ with respect to $n$ have been reduced to the derivatives of $\Delta$ with respect to $n$, because these quantities can be calculated directly from Eq.~(\ref{eq:deltafromdensity}).

Eq.~(\ref{eq:expDensSwing0}) is the analogous of Eq.~(\ref{eq:swingtdens}) for a homogeneous state and, as discussed in Sec.~\ref{sect:systemsolution}, can be used together with the other equations in ${\cal E}$, trivially expanded at zero order, to find the equilibrium homogeneous value of the variables in ${\cal V}$.

Eq.~(\ref{eq:expDensSwing2}) shows that a finite oscillating fluctuation $n^{(1)}(x,t)$ can feed into dc fluctuations $\delta n(x)$, $\delta U_{\rm t}(x)$.
This effect, called ``rectification,'' is generic of systems with a nonlinear ``characteristic curve,'' as for example for a FET.
In our case, the characteristic curves correspond to the relations between $U_{\rm t}(n)$ or $\Delta(n)$ and $n$, shown in Figs.~\ref{fig:electrostaticsthick} and~\ref{fig:electrostaticsgate}.
While in the rest of this paper we will focus on the rectification generated by the \emph{hydrodynamic nonlinearities} due to density fluctuations at \emph{finite} wavelength, (not merely due to the nonlinearity of the characteristic curves,) it is important to keep in mind that the rectification process described by Eq.~(\ref{eq:expDensSwing2}) is present in a homogeneous state as well.

Finally, it is useful to calculate the derivatives of the total gate to channel swing $U$ defined in Eq.~(\ref{eq:swings}).
Deriving Eq.~(\ref{eq:swingb}) with respect to $n$, we immediately find
\begin{equation}\label{eq:effcap}
\bar{U}' = 2 \bar{U}_{\rm t}' - \frac{\bar{\Delta}'}{e} = \frac{e}{(C_{\rm t} + C_{\rm t})} \left ( 2 + \frac{C_{\rm b} - C_{\rm t}}{e} \frac{\bar{\Delta}'}{e} \right ) \equiv \frac{e}{C}~,
\end{equation}
\begin{equation}\label{eq:quantumcap}
\bar{U}'' = 2 \bar{U}_{\rm t}'' - \frac{\bar{\Delta}'}{e} = \frac{C_{\rm b} - C_{\rm t}}{C_{\rm t} + C_{\rm b}} \frac{\bar{\Delta}''}{e}~.
\end{equation}
The quantity $C$ introduced in Eq.~(\ref{eq:effcap}) is an effective capacitance per unit area, which relates $U$ and $n$ as if they were the gate to channel swing and the carrier density in a standard single-gate, single-layer setup.
The asymmetry between the role played by the top and bottom capacitances in Eq.~(\ref{eq:effcap}) is simply due to the definition of $\Delta$.
Indeed, if one exchanges the top and bottom indices in Eq.~(\ref{eq:deltadefinition}) and in Eq.~(\ref{eq:effcap}), the definition of $C$ is unchanged.
We also note that $C$ may diverge for particular choices of the capacitances.
In this case, the linearization procedure discussed in this section cannot be used (because a small fluctuation of the swing produces a divergent fluctuation of the density) and one has to resort to the nonlinear solution of the system ${\cal E}$ as discussed in Sec.~\ref{sect:systemsolution}.
In the limit $\Delta = 0$, the effective capacitance reduces to the average $C = (C_{\rm t} + C_{\rm b}) / 2$ of the top and bottom capacitances.

\section{Hydrodynamic equations}
\label{sec:hydrodynamicequations}

\subsection{Continuity and Euler equations}

We describe the time-evolution of the electron system using electron hydrodynamic equations.~\cite{tomadin_prl_2014, torre_prb_2015, narozhny_prb_2015, levitov_natphys_2016, narozhny_annphys_2017, guo_pnas_2017, gabbana_prl_2018, shytov_prl_2018, ledwith_prl_2019}
Here, we neglect shear and bulk viscosities,~\cite{principi_prb_2016} consistently with the hypothesis~(\ref{eq:fermistep}) of low temperature.
The hydrodynamic variables are the local drift velocity $v(x,t)$ and the total density $n(x,t)$ on the bilayer.
We remind the reader that $v(x,t)$ is the parameter entering the quasi-equilibrium distribution (\ref{eq:distrohydro}) and $n(x,t)$ is defined as in (\ref{eq:totaldensity}).
The particle current is given by
\begin{equation}\label{eq:defcurrent}
j(x,t) = n(x,t) v(x,t)~.
\end{equation}

The first hydrodynamic equation is the continuity equation, which originates from conservation of particle number.
The continuity equation reads
\begin{equation}
\label{eq:continuity}
\partial_{t} n(x,t) + \partial_{x} \lbrack n(x,t) v(x,t) \rbrack = 0~.
\end{equation}
The second hydrodynamic equation is the Euler equation, which originates from conservation of momentum.
The Euler equation is derived from the Boltzmann equation, where the collisional integral vanishes due to the choice~(\ref{eq:distrohydro}) for the probability distribution.
In the case of single-layer graphene, a detailed derivation has been reported in Ref.~\cite{tomadin_prb_2013}, where it is emphasized the crucial role of the {\it nonlinear} relation between the average carrier momentum and the average carrier velocity.
Due to this nonlinear relation, the Euler equation is amenable of an analytical treatment in the limit $v(x,t) \ll v_{\rm F}$ of small drift velocities only.
In the case of the BLG, we proceed along the lines of the derivation detailed in Ref.~\cite{tomadin_prb_2013} and, in the limit $v(x,t) \ll v_{\rm F}$, we obtain the following Euler equation
\begin{eqnarray}
\label{eq:euler_full}
\gamma(x,t) \lbrack \partial_t v(x,t) + v(x,t) \partial_x v(x,t) \rbrack & &  \nonumber \\
+ \frac{e}{m} \frac{n_{{\rm e}}(x,t) - n_{{\rm h}}(x,t)}{n(x,t)} \partial_x U (x,t) & & \nonumber \\
+ \frac{1}{m n(x,t)} \partial_x P(x,t) + v(x,t) \partial_t \gamma(x,t) & & \nonumber \\
+ v(x,t)^{2} \partial_x \gamma(x,t) + \frac{1}{\tau} \gamma(x,t) v(x,t) & = & 0~, \nonumber \\
\end{eqnarray}
where $U(x,t)$ is defined as in Eq.~(\ref{eq:swings}) and the variable $P(x,t)$ is the pressure of the electron liquid.
The last term represents friction and $\tau$ is a phenomenological relaxation time.
In the limit $k_{\rm B}T \ll \varepsilon_{\rm F}$, the dimensionless coefficient $\gamma(x,t)$ reads:
\begin{equation}\label{eq:presschange}
\gamma(x,t) = \sqrt{1 + \left(\frac{m \Delta(x,t)}{\hbar^{2} \pi n(x,t)}\right)^2}~.
\end{equation}
This term represent the local relative change in pressure of the electronic fluid due to the existence of a non-zero inter-layer potential energy difference $\Delta(x,t)$.
If we take the limit $\Delta(x,t) = 0$, then $\gamma(x,t) = 1$ and we recover the standard Euler equation for two parabolic carriers with opposite charge.
[In this limit, however, the two-bands model~(\ref{eq:blghamil}) is not justified.]

It is a good approximation to neglect the gradient of the pressure $\partial_{x} P(x,t)$ with respect to the Coulomb force~\cite{tomadin_prb_2013}.
Finally, for definiteness, we assume that the Fermi energy lies in the conduction band, so that $n_{{\rm e}}(x,t) \simeq n(x,t) > 0$.
In this case, the Euler equation simplifies to:
\begin{eqnarray}
\label{eq:euler}
& & \gamma(x,t) \left[ \partial_t v(x,t) + v(x,t) \partial_x v(x,t)\right] = -\frac{e}{m}\partial_x U(x,t) \nonumber \\
& & - v(x,t) \partial_t \gamma(x,t) - v(x,t)^{2} \partial_x \gamma(x,t)- \frac{1}{\tau} \gamma(x,t) v(x,t)~. \nonumber \\
\end{eqnarray}
This is the form of the Euler equation that we use in the following sections to study the photoresponse of a graphene bilayer in the Dyakonov-Shur scheme.
The quantities $\Delta(x,t)$, $\gamma(x,t)$, and $U(x,t)$ are functions of $n(x,t)$, which can be computed following the procedure explained in Secs.~\ref{sect:systemsolution} and~\ref{sect:elstaticslinear}.
Hence, the continuity and Euler equations~(\ref{eq:continuity}) and~(\ref{eq:euler}) define a system of two coupled differential equations for the two variables $n(x,t)$ and $v(x,t)$.

\subsection{Boundary conditions}
\label{sect:boundaryconditions}

As anticipated in Sec.~\ref{sect:introduction}, our aim is to calculate the dc voltage difference between the drain and the source contacts, which arises when an oscillating voltage difference is applied between the top gate and the top graphene layer at the source contact.
We impose that the drain contact is floating and no current flows.
This setup is represented by the following BCs~\cite{dyakonov_ieee_1996a}:
\begin{equation}\label{eq:bcsource}
U_{\rm t}(x=0,t) = U_{0} + U_{\rm a} \cos(\omega t)~,
\end{equation}
\begin{equation}\label{eq:bcdrain}
j(x=L,t) = 0~.
\end{equation}

To solve Eqs.~(\ref{eq:continuity}) and~(\ref{eq:euler}) with the BCs~(\ref{eq:bcsource}) and~(\ref{eq:bcdrain}), it is convenient to resort to the second-order expansion introduced in Eq.~(\ref{eq:expansion2o}), which includes both oscillating and dc components.
Explicitly, the expansion reads:
\begin{equation}\label{eq:expdensity}
n(x,t) = \bar{n} + n^{(1)}(x,t) + \delta n(x) + n^{(2)}(x,t)~,
\end{equation}
\begin{equation}\label{eq:expvelocity}
v(x,t) = \bar{v} + v^{(1)}(x,t) + \delta v (x) + v^{(2)}(x,t)~.
\end{equation}
Here, $\bar{n}$ and $\bar{v}$ is the steady uniform solution, compatible with the BCs.
The linear response of the system to the driving is represented by $n^{(1)}(x,t)$ and $v^{(1)}(x,t)$ and takes the form of a linear combinations of plasma waves, which will be calculated in Sec.~\ref{sec:plasmawaves}, with amplitude proportional to $U_{\rm a}$.
The remaining terms represent density and velocity fluctuations with amplitude proportional to $U_{\rm a}^{2}$, which arise due to the nonlinear terms in Eqs.~(\ref{eq:continuity}) and~(\ref{eq:euler}).

The components of the expansion of the current $j(x,t)$ can be obtained from the components of $n(x,t)$ and $v(x,t)$ in Eqs.~(\ref{eq:expdensity}) and~(\ref{eq:expvelocity}).
Expanding the left- and right-hand side of the definition~(\ref{eq:defcurrent}), and equating components of different order and frequency, we find the relations
\begin{equation}\label{eq:expcurrent0}
\bar{j} = \bar{n} \bar{v}~,
\end{equation}
\begin{equation}\label{eq:expcurrent1}
j^{(1)}(x,t) = \bar{n} v^{(1)}(x,t) + \bar{v} n^{(1)}(x,t)~,
\end{equation}
\begin{equation}\label{eq:expcurrent2}
\delta j(x) = \bar{n} \delta v(x) + \left \langle n^{(1)}(x,t) v^{(1)}(x,t) \right \rangle_{t}~,
\end{equation}
where the time-average is defined in Eq.~(\ref{eq:timeaverage}).

From Eq.~(\ref{eq:bcsource}) we obtain the BCs for the components of the density at $x = 0$, by setting $\bar{U}_{\rm t} = 0$, $U_{\rm t}^{(1)}(0,t) = U_{\rm a} \cos{(\omega t)}$, and $\delta U_{\rm t}(0) = 0$ in Eqs.~(\ref{eq:expDensSwing0})--(\ref{eq:expDensSwing2}).
The value of $U_{0}$, together with the parameter $V_{\rm t} - V_{\rm b}$, determines the equilibrium density $\bar{n} = n(U_{0})$, as seen from Eqs.~(\ref{eq:swingtdens}) and~(\ref{eq:deltafromdensity}).
Moreover, we find that
\begin{equation}\label{eq:denssource}
n^{(1)}(0,t) = \frac{U_{\rm a}}{\bar{U}_{\rm t}'} \cos{(\omega t)}~.
\end{equation}

Eq.~(\ref{eq:bcdrain}) implies that each component in the current expansion has to be set to zero at $x = L$.
Setting the right-hand sides of Eqs.~(\ref{eq:expcurrent0})--(\ref{eq:expcurrent2}) equal to zero we obtain the BCs for the components of the velocity at $x = L$: $\bar{v} = 0$,
\begin{equation}\label{eq:veldrain}
v^{(1)}(L,t) = 0~,
\end{equation}
and $\bar{n} \delta v(L) = - \left \langle n^{(1)}(L,t) v^{(1)}(L,t) \right \rangle_{t}$.
The six BCs for the components of the density and of the velocity will be used in the following algebraic manipulations.

\section{Plasma waves}
\label{sec:plasmawaves}

\subsection{Freely-propagating plasma waves}

It is well known that the gated electron system in the hydrodynamic regime supports the propagation of longitudinal modes known as plasma waves,~\cite{dyakonov_prl_1993,dyakonov_prb_1995,dyakonov_ieee_1996a,dyakonov_ieee_1996b} with linear dispersion $\omega = s k$, where $\omega$, $s$, and $k$ are the angular frequency, the speed, and the wave vector of the plasma wave, respectively.
Before we proceed to the calculation of the DS photoresponse, we calculate the plasma-wave speed $s$.
In systems with a parabolic electron dispersion, the average value $\bar{v}$ of the electron fluid speed affects the plasma-wave speed $s$ consistently with the Galilean invariance underlying the electronic spectrum, i.e.~$\omega = (\bar{v} \pm s)k$~\cite{dyakonov_prl_1993}.
Notably, it has been shown that this is not the case in graphene~\cite{tomadin_prb_2013} where a measurement of the plasma-wave dispersion would reveal the absence of a Galilean-invariant spectrum.~\cite{polini_macdonald_vignale}
When particular boundary conditions (BCs) are imposed in a finite-size sample, the Galilean invariance is of course broken and the plasma-wave have a BCs-dependent spectrum which may even feature unstable modes giving rise to self-sustained oscillations.~\cite{dyakonov_prl_1993}

In our case, as discussed in Sec.~\ref{sect:boundaryconditions}, a bias current through the sample is absent, thus we calculate the plasma-wave speed assuming $\bar{v} = 0$.
Expanding the continuity and Euler equation at first order we find:
\begin{equation}\label{eq:continuity_linear}
\partial_t n^{(1)}(x,t) + \bar{n} \partial_x v^{(1)}(x,t)=0~,
\end{equation}
\begin{equation}\label{eq:euler_linear}
\bar{n} \partial_t v^{(1)}(x,t) = \frac{e^{2} \bar{n}}{m C \bar{\gamma}} \partial_x n^{(1)}(x,t) - \frac{1}{\tau} \bar{n}  v^{(1)}(x,t)~,
\end{equation}
where we use the notation introduced in Eq.~(\ref{eq:notationderiv}), the effective capacitance $C$ defined in Eq.~(\ref{eq:effcap}), and the first-order expansion of density and of the velocity introduced in Eqs.~(\ref{eq:expdensity}) and ~(\ref{eq:expvelocity}).
To solve Eqs.~(\ref{eq:continuity_linear}) and (\ref{eq:euler_linear}), we use an ansatz representing traveling waves:
\begin{eqnarray}\label{eq:planemodes}
n^{(1)}(x,t) & = & n_{K} e^{i K x - i \Omega t} + n_{K}^* e^{- i K^{\ast} x + i \Omega t}, \nonumber \\
v^{(1)}(x,t) & = & v_{K} e^{i K x - i \Omega t } + v_{K}^* e^{ - i K^{\ast} x + i \Omega t}~,
\end{eqnarray}
where we assume that the wave vector $K$ is in general complex and the frequency $\Omega$ is real.
We require that ${\rm Re}[K] \times {\rm Im}[K] > 0$, i.e.~the traveling wave is damped in the direction of propagation.

We note that, usually, the calculation of the spectrum assumes that the wave vector is real and the frequency is in general complex.
Which choice is most convenient, depends on the BCs.
The usual choice is appropriate when the BCs are local in time but respect the spatial symmetry of the system, i.e.~they represent the initial value of the amplitude of a well-defined spatial mode.
The spatial quantum number is conserved in the time-evolution but the mode amplitude decays in time.
Our choice, instead, is more appropriate for BCs such as~(\ref{eq:bcsource}) and (\ref{eq:bcdrain}), which are local in space and break the translational invariance of the continuity and Euler equations, but oscillate in time with a given frequency.
The mode amplitude is then forced to oscillate periodically, but it decays in space away from the point where the driving is applied.

Inserting the ansatz~(\ref{eq:planemodes}) into Eqs.~(\ref{eq:continuity_linear}) and~(\ref{eq:euler_linear}), we obtain the following linear system for the coefficients $v_{K}$, $n_{K}$:
\begin{eqnarray} \label{eq:sysdispersion}
\left( \begin{array}{cc}
\Omega + i/\tau & - K e^{2} / (m C \bar{\gamma}) \\
- \bar{n} K & \Omega
\end{array}
\right)
\left (
\begin{array}{c}
v_{K} \\
n_{K}
\end{array} \right ) = 0~.
\end{eqnarray}
Solving the associated secular equation we find that there are two modes at frequency $\Omega$, with wave vectors $K = k(\Omega)$ and $K = -k(\Omega)$, where we introduce the dispersion relation
\begin{equation}\label{eq:dispersion}
k(\omega) \equiv \frac{\omega}{s} \sqrt{1 + \frac{i}{\omega \tau}}, \quad
s \equiv \sqrt{\frac{e^{2} \bar{n}}{m C \bar{\gamma}}}~,
\end{equation}
and the plasma-wave speed $s$.
If $\omega > 0$, the complex square root  is chosen with a positive imaginary part, to fulfill the condition ${\rm Re}[K] \times {\rm Im}[K] > 0$.

In the limit $\omega \tau \gg 1$ of negligible friction, we recover the linear plasma-wave dispersion relation.
The expression of the plasma-wave speed valid in the parabolic case~\cite{dyakonov_prl_1993} is recovered in the limit $\Delta(x,t) = 0$, when $\bar{\gamma} = 1$ and $C = (C_{\rm t} + C_{\rm b}) / 2$.
The second line of the linear system~(\ref{eq:sysdispersion}) gives the relation between the coefficients of the velocity and the density in the ansatz
\begin{equation}\label{eq:reldensvellin}
v_{K} = \frac{\Omega}{K} \frac{n_{K}}{\bar{n}}~,
\end{equation}
which is just a reformulation of the continuity equation.

\subsection{Plasma-waves in the finite-size system}

A generic solution of the hydrodynamic equations, fulfilling the BCs defined in Sec.~\ref{sect:boundaryconditions}, is necessarily a linear superposition of the two modes with $\Omega = \omega$ and $K = \pm k(\omega)$:
\begin{eqnarray}\label{eq:modeswithbc}
n^{(1)}(x,t) & = & n_{k(\omega)} e^{i k(\omega) x - i \omega t} + n_{-k(\omega)} e^{-i k(\omega) x - i \omega t} + \mbox{c.c.}, \nonumber \\
v^{(1)}(x,t) & = & v_{k(\omega)} e^{i k(\omega) x - i \omega t } + v_{-k(\omega)} e^{-i k(\omega) x - i \omega t } + \mbox{c.c.}~. \nonumber \\
\end{eqnarray}
In other words, traveling waves are reflected at the boundaries of the system, such that the solution is a superposition of traveling waves with opposite wave vectors.
Each traveling wave is damped in space in the direction of propagation.

Inserting these modes superposition into Eq.~(\ref{eq:denssource}) we find $n_{k(\omega)} + n_{-k(\omega)} = n_{\rm a} / 2$, and into Eq.~(\ref{eq:veldrain}) we find $v_{k(\omega)} e^{i k(\omega) L} + v_{-k(\omega)} e^{-i k(\omega) L} = 0$.
Using Eq.~(\ref{eq:reldensvellin}), we obtain the coefficients
\begin{eqnarray}
n_{\pm k(\omega)} & = & \frac{U_{\rm a}}{2 \bar{U}_{\rm t}'} \frac{1}{1 + e^{\pm 2 i k(\omega) L}}, \nonumber \\
v_{\pm k(\omega)} & = & \pm \frac{U_{\rm a}}{2 \bar{U}_{\rm t}' \bar{n}} \frac{\omega}{k(\omega)} \frac{1}{1 + e^{\pm 2 i k(\omega) L}}~. \nonumber \\
\end{eqnarray}

\section{Photovoltage}
\label{sect:photovoltage}

\subsection{Expression for the photovoltage from the hydrodynamic equations}

The photovoltage is the difference of the dc gate to channel swing for the top layer
\begin{equation}\label{eq:photovoltdef}
\Delta U_{\rm t} \equiv \delta U_{\rm t}(L) - \delta U_{\rm t}(0)
\end{equation}
between the drain and the source contacts.
Because of the BC~(\ref{eq:bcsource}), the expression of the photovoltage simplifies to $\Delta U_{\rm t} = \delta U_{\rm t}(L)$.
In terms of hydrodynamic variables, with Eq.~(\ref{eq:expDensSwing2}) we find
\begin{equation}\label{eq:photovolt01}
\Delta U_{\rm t} = \bar{U}_{\rm t}' \delta n(L) + \frac{1}{2} \bar{U}_{\rm t}'' \langle n^{(1)}(L,t)^{2} \rangle_{t}~.
\end{equation}
The time-average on the right-hand side can be directly calculated using the expressions~(\ref{eq:modeswithbc}) of the linear modes.
Now we proceed to express the quantity $\delta n(L)$ in terms of averages of linear modes as well.
We first expand Eqs.~(\ref{eq:continuity}) and~(\ref{eq:euler}) to second order, obtaining:
\begin{eqnarray}
\partial_t n^{(2)}(x,t) + \bar{n}\partial_x \lbrack \delta v(x) + v^{(2)}(x,t)\rbrack + & & \nonumber \\
\partial_x \lbrack n^{(1)}(x,t)v^{(1)}(x,t)\rbrack & = & 0~,
\end{eqnarray}
and
\begin{eqnarray}
& & \bar{n} \left \lbrack \partial_{t} v^{(2)}(x,t) + v^{(1)}(x,t) \partial_{x} v^{(1)}(x,t) \right \rbrack = \nonumber \\
& & - s^{2}
\partial_{x} \lbrack \delta n(x) + n^{(2)}(x,t)\rbrack - \frac{1}{\tau} \bar{n} \left \lbrack \delta v(x) + v^{(2)}(x,t) \right \rbrack \nonumber \\
& & - s^{2} \frac{\bar{U}''}{\bar{U}'} n^{(1)}(x,t) \partial_{x} n^{(1)}(x,t) \nonumber \\
& & - \frac{\bar{\gamma}' \bar{n}}{\bar{\gamma}}
\left ( \partial_{t} - \frac{1}{\tau} \right ) \left \lbrack n^{(1)}(x,t) v^{(1)}(x,t) \right \rbrack~.
\end{eqnarray}

The derivative of the function $\gamma$ with respect to the density can be written as
\begin{equation}\label{eq:gammader}
\bar{\gamma}' = \frac{\bar{\gamma}^{2} - 1}{\bar{\gamma} \bar{n}} \left (\frac{\bar{\Delta}' \bar{n}}{\bar{\Delta}}  - 1\right )
\sim \left ( \frac{m}{\hbar^{2} \pi \bar{n}} \right )^{2} \bar{\Delta} \left ( \bar{\Delta}' \bar{n} - \bar{\Delta} \right )~.
\end{equation}
The asymptotic form is valid for $\bar{\Delta} \ll \hbar^{2} \pi \bar{n} / m$ and shows that in this limit $\bar{\gamma}'$ vanishes.
Then we take the time-average of the second-order equations over one period of the applied voltage, obtaining:
\begin{equation}\label{eq:conttimeav}
\bar{n} \partial_{x} \delta v(x)  + \partial_{x} \langle n^{(1)}(x,t) v^{(1)}(x,t) \rangle_{t} = 0~,
\end{equation}
and
\begin{eqnarray}\label{eq:eulertimeav}
& & \bar{n} \langle v^{(1)}(x,t) \partial_{x} v^{(1)}(x,t)\rangle_{t} = -s^{2} \partial_{x} \delta n(x) - \frac{1}{\tau} \bar{n} \delta v(x) - \nonumber \\
& & s^{2} \frac{\bar{U}''}{\bar{U}'} \langle  n^{(1)}(x,t) \partial_{x} n^{(1)}(x,t) \rangle_{t}
-\frac{1}{\tau} \frac{\bar{\gamma}' \bar{n}}
{\bar{\gamma}} \langle n^{(1)}(x,t) v^{(1)}(x,t) \rangle_{t}~. \nonumber \\
\end{eqnarray}
We now integrate Eq.~(\ref{eq:conttimeav}) in space from a generic $x$ to $x = L$ and substitute the result for $\delta v(x)$ into Eq.~(\ref{eq:eulertimeav}).
We then integrate the resulting equation in space from $x = 0$ to $x = L$ and we find the desired expression for $\delta n(L)$:
\begin{eqnarray}
\delta n(L) & = & \delta n(0) +
\left ( 1 - \frac{\bar{\gamma}' \bar{n}}{\bar{\gamma}} \right ) \frac{1}{s^{2} \tau} \times \nonumber \\
& & \int_{0}^{L} dx \left \langle n^{(1)}(x,t) v^{(1)}(x,t) \right \rangle_{t} - \nonumber \\
& & \frac{1}{2} \frac{\bar{U}''}{\bar{U}'} \left \lbrack \left \langle n^{(1)}(L,t)^{2} \right \rangle_{t} - \left \langle n^{(1)}(0,t)^{2} \right \rangle_{t} \right \rbrack + \nonumber \\
& & s\frac{1}{2} \frac{\bar{n}}{s^{2}} \langle v^{(1)}(0,t)^{2} \rangle_{t}~.
\end{eqnarray}

\begin{figure}
(a)\includegraphics[width=0.75\linewidth]{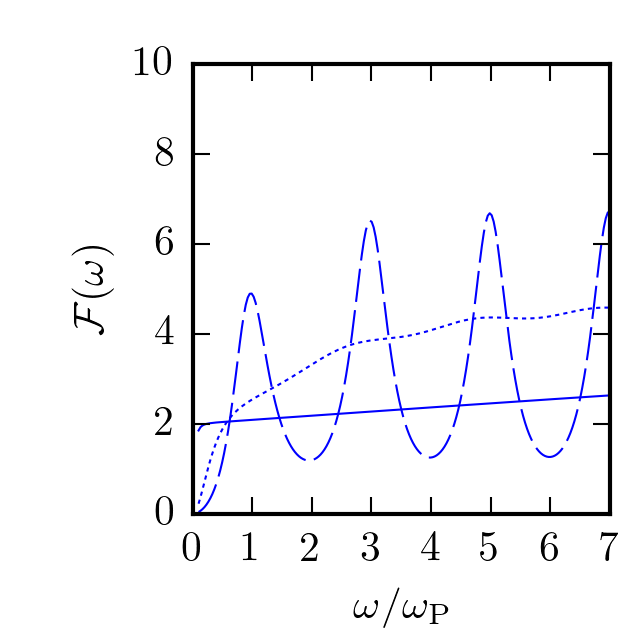}
(b)\includegraphics[width=0.75\linewidth]{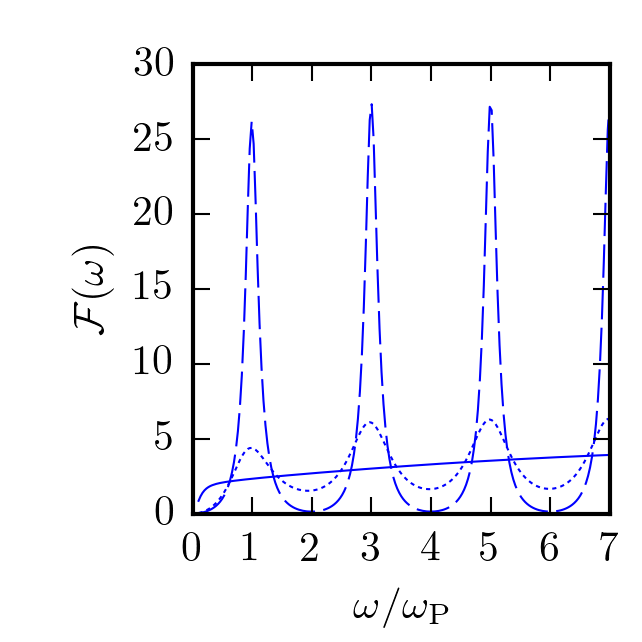}
\caption{\label{fig:photoresponse}
Photoresponse function ${\cal F}(\omega)$ as a function of the ratio between the angular frequency $\omega$ of the applied voltage and the fundamental plasma angular frequency $\omega_{\rm P}$ defined in Eq.~(\ref{eq:fundfreq}), for (a) $d_{\rm t} = 30~{\rm nm}$ and (b) $d_{\rm t} = 300~{\rm nm}$.
The other parameters are: $d_{\rm b} = 100~{\rm nm}$, $\tau = 1~{\rm ps}$, $L = 5~\mu{\rm m}$, $V_{\rm t} - V_{\rm b} = 100~{\rm V}$, $n = 0.1$ (solid), $1.0$ (dotted), and $5.0\times 10^{12}~{\rm cm}^{-2}$ (dashed). }
\end{figure}

Inserting this expression into~(\ref{eq:photovolt01}) we obtain the photovoltage in terms of average of linear modes
\begin{eqnarray}\label{eq:photovoltaverages}
\Delta U_{\rm t} & = &
\left ( 1 - \frac{\bar{\gamma}' \bar{n}}{\bar{\gamma}} \right ) \frac{\bar{U}_{\rm t}'}{s^{2} \tau} \int_{0}^{L} dx \left \langle n^{(1)}(x,t) v^{(1)}(x,t) \right \rangle_{t} \nonumber + \\
& & \frac{1}{2} \frac{\bar{U}_{\rm t}' \bar{n}}{s^{2}} \langle v^{(1)}(0,t)^{2} \rangle_{t} +
 \nonumber \\
& & \frac{1}{2} \left ( \bar{U}_{\rm t}'' - \bar{U}_{\rm t}' \frac{\bar{U}''}{\bar{U}'} \right ) \left \lbrack \left \langle n^{(1)}(L,t)^{2} \right \rangle_{t} - \left \langle n^{(1)}(0,t)^{2} \right \rangle_{t} \right \rbrack~. \nonumber \\
\end{eqnarray}
In this expression, the terms proportional the first derivative of the function $\gamma$ or to the second derivatives of the swings are peculiar to the bilayer system considered here, and do not appear in the analogous expression for a 2DEG channel.
The term $\bar{\gamma}'$ represents the change of pressure with density and $\bar{U}''$, $\bar{U}_{\rm t}''$ represent \emph{quantum capacitance} effects, i.e.~the nonlinear scaling of the electric potentials with the electric charge on the conducting surfaces.
Both these effects are due to the existence of the asymmetry $\Delta(x,t)$ between the states in the two layers, generated by asymmetric gating of the top and bottom layer.
These derivatives are expressed in terms of derivatives of $\Delta(n)$ with respect to $n$ at the equilibrium density $n = \bar{n}$ in Eqs.~(\ref{eq:expDensSwing1}), (\ref{eq:expDensSwing2}), (\ref{eq:effcap}), (\ref{eq:quantumcap}), and (\ref{eq:gammader}).

\begin{figure}
(a)\includegraphics[width=0.75\linewidth]{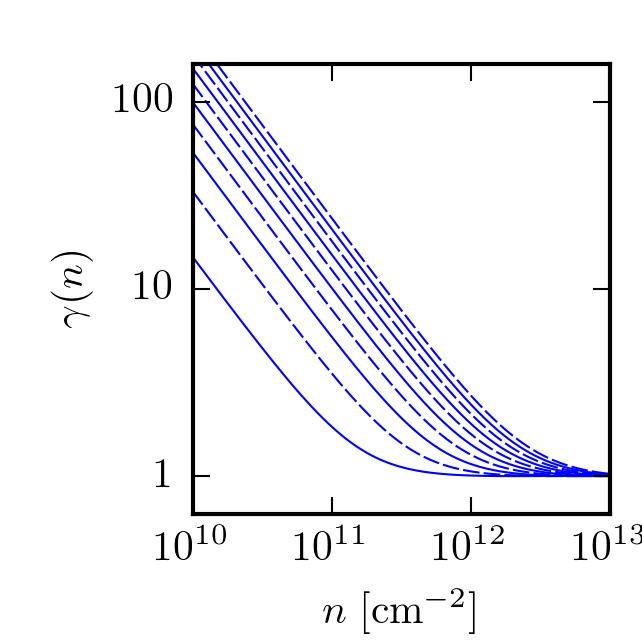}
(b)\includegraphics[width=0.75\linewidth]{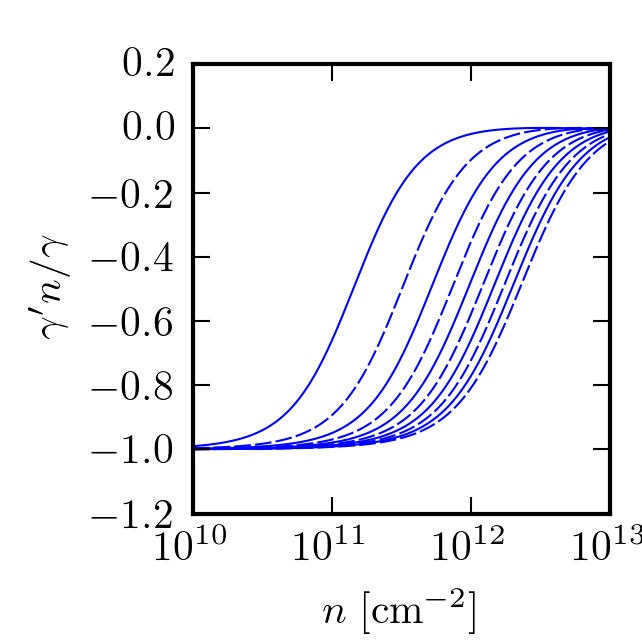}
\caption{\label{fig:gammafunction}
(a) The function $\gamma(n)$ and (b) its derivative $\gamma'$, rescaled to be dimensionless, as a function of the density $n$.
Different lines correspond to values of $V_{\rm t} - V_{\rm b}$ equally spaced from $V_{\rm t} - V_{\rm b} = 10~{\rm V}$ (solid) to $V_{\rm t} - V_{\rm b} = 100~{\rm V}$ (dashed).
The other parameters are: $d_{\rm t} = 300~{\rm nm}$, $d_{\rm b} = 100~{\rm nm}$, $\tau = 1~{\rm ps}$, $L = 5~\mu{\rm m}$. }
\end{figure}

\begin{figure}
(a)\includegraphics[width=0.75\linewidth]{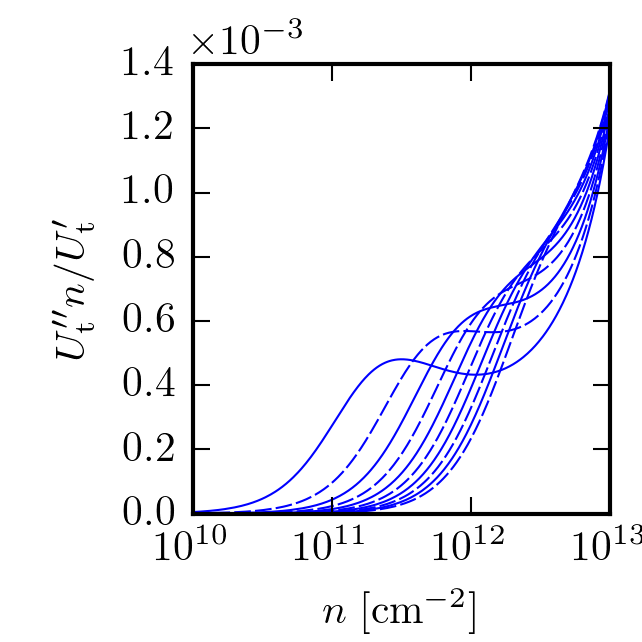}
(b)\includegraphics[width=0.75\linewidth]{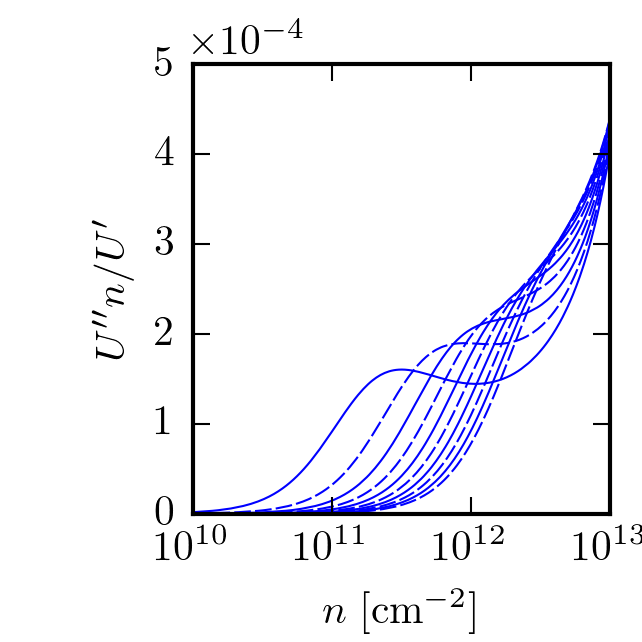}
\caption{\label{fig:swings2}
(a) Second derivative of the gate to channel swing for the top gate $U_{\rm t}(n)$ and (b) of the sum of the swings $U(n)$, as a function of the density $n$.
Parameters are chosen as in Fig.~\ref{fig:gammafunction}. }
\end{figure}

\subsection{Photoresponse function}

The values of the space- and time-averages in Eq.~(\ref{eq:photovoltaverages}) can be evaluated directly using the expression~(\ref{eq:modeswithbc}) of the linear modes.
We find
\begin{equation}\label{eq:photovoltage}
\Delta U_{\rm t} = \frac{1}{4} \frac{U_{\rm a}^{2}}{\bar{U}_{\rm t}' \bar{n}} ~ {\cal F}(\omega)~,
\end{equation}
which is explicitly a second-order expression in the strength $U_{\rm a}$ of the driving field.
The frequency dependence is given by the dimensionless nonlinear response function
\begin{eqnarray}\label{eq:responsefunction}
{\cal F}(\omega) & = &
\left ( 1 - \frac{\bar{\gamma}' \bar{n}}{\bar{\gamma}} - \frac{\bar{U}_{\rm t}'' \bar{n}}{\bar{U}_{\rm t}'} + \frac{\bar{U}'' \bar{n}}{\bar{U}'} \right ) \times \nonumber \\
& & \left \lbrack 1 - \frac{1}{\cos{(k(\omega) L)} \cos{(k(\omega)^{\ast} L)}} \right \rbrack + \nonumber \\
& & \left ( 1 - \frac{1}{2} \frac{\bar{\gamma}' \bar{n}}{\bar{\gamma}} \right ) \beta(\omega) \tan{(k(\omega) L)} \tan{(k(\omega)^{\ast} L)}~, \nonumber \\
\end{eqnarray}
where the function $k(\omega)$ is given in Eq.~(\ref{eq:dispersion}) and
\begin{equation}\label{eq:betafunction}
\beta(\omega) \equiv \frac{2 \omega \tau}{\sqrt{1 + (\omega \tau)^2}}~.
\end{equation}

Eq.~(\ref{eq:responsefunction}) is the main result of this work.
The response function is shown in Fig.~\ref{fig:photoresponse} for a few sets of parameters.
The typical frequency dependence of the photoresponse function consists of resonant maxima at the odd multiples of the fundamental plasma angular frequency
\begin{equation}\label{eq:fundfreq}
\omega_{\rm P} = \frac{1}{4} \times 2 \pi \frac{s}{L}~.
\end{equation}
The ratio $s / (2 L)$ corresponds to the frequency  arising when a wave travels along the system of length $L$ with speed $s$ and is reflected by the boundaries.
The extra factor $1 / 2$ is due to the asymmetric BCs imposed on the system.
The frequency dependence and the scaling with the momentum relaxation time $\tau$ has already been studied in detail in the literature.~\cite{dyakonov_prl_1993,dyakonov_prb_1995,dyakonov_ieee_1996a,dyakonov_ieee_1996b}

We point out that $k(\omega)$ depends on $\bar{n}$ through $s$, see Eq.~(\ref{eq:dispersion}), such that all terms in Eq.~(\ref{eq:responsefunction}) depend on the equilibrium density.
However, when $\omega \tau \gg 1$, $\beta \to 2$ and the periodic functions in Eq.~(\ref{eq:responsefunction}) depend only on the ratio $\omega / \omega_{\rm P}$.
In other words, changing the density merely rescales the response function in frequency space.
In this regime, \emph{the amplitude of the photoresponse is governed by the coefficients} $\gamma$, $U_{\rm t}$, and their derivatives with respect to the density, which are shown in Figs.~\ref{fig:gammafunction} and~\ref{fig:swings2} for a set of parameters.
We notice, in particular that the density-dependence of the relative pressure change $\gamma$ spans several orders of magnitudes, while the quantum capacitance terms are non-monotonic.
The final combination of these terms in the prefactor of Eq.~(\ref{eq:responsefunction}) depends on the choice of the parameters. 
Thus, compared to the DS photoresponse of a single-layer graphene FET, Eq.~(\ref{eq:responsefunction}) displays a more varied dependence on the system's parameters, which translates into an easier tuning of the FET's operation point to a sweet spot by electrical doping.

\section{Summary and perspectives}
\label{sec:summary}

In this paper, we have presented a detailed theory of the photoresponse in a dual-gated, bilayer-graphene field-effect transitor, based on the effect of nonlinear interference of plasma-waves introduced by Dyakonov and Shur.

The nonlinear relations between the electron density and the other quantum, thermodynamic, and electrostatic variables are particularly intricated in this geometry and lead to the effective constitutive equation (\ref{eq:deltafromdensity}), which describes the coupling between the macroscopic electron density $n$ and the microscopic asymmetry potential $\Delta$.
The hydrodynamic equations describing the electron system include a new term (\ref{eq:presschange}) which represents the effect of the asymmetry potential on the electron pressure.
As a consequence, the physics of the plasma waves is considerably more involved than in single-layer graphene, because a local oscillation of the electron density induces nonlinear oscillations of the asymmetry potential.
This leads, in turn, to non-trivial quantum capacitance effects which appear as leading order corrections to the expression for the photoresponse (\ref{eq:responsefunction}).
The final photoresponse at fixed frequency varies on a broad range, depending on the system's parameters, making it easier to tune a photodetector to an operational sweet spot by changing the equilibrium density.

Recent experimental results~\cite{bandurin_natcommun_2018}, demonstrating resonant photodetection using bilayer graphene encapsulated in hexagonal boron nitride, show the possibility to verify our predictions in real-world devices, with the aim of maximizing the photoresponse by carefully tuning the system's parameters.

More generally, we anticipate that the investigation of other two-dimensional van der Waals heterostructures might lead to an enhanced platform for Dyakonov-Shur photodetection.
Indeed, in recent years, a large number of two-dimensional materials has been explored;~\cite{avouris_heinz_low_2017, materials_topical_review}
the propagation of coupled light-matter excitations in these materials has been the subject of continuing investigations;~\cite{tielrooij_natnanotech_2018, sunku_science_2018, epstein_2dmater_2020} and imaging techniques for two-dimensional samples in the terahertz range have been refined.~\cite{mitrofanov_scirep_2017,giordano_optica_2018}
Electronic states in these materials are described in terms of microscopic Hamiltonians which, when appropriately taken into account in the derivation of the photoresponse, might lead to enhanced quantum capacitance effects, or even more exotic couplings between quantum and thermodynamic degrees of freedom, improving the photoresponse.

\appendix

\section{Broadband photoresponse}

It is of practical interest to evaluate the photovoltage $\Delta U_{\rm t}$, see Eq.~(\ref{eq:photovoltage}), in the limit (i) $\omega \tau \ll 1$ or broadband limit, corresponding to overdamped plasma oscillations, and (ii) $L \gg \omega s$, corresponding to a long detection device.
These limits cover the majority of currently available devices where coherent propagation and interference of plasma waves over length scales of several microns is hampered by losses.
In these limits, the functional dependence of the photovoltage on the frequency in Eq.~(\ref{eq:responsefunction}) vanishes and we find
\begin{equation}\label{eq:dubroadband}
\Delta U_{\rm t} = \frac{1}{4} \frac{U_{\rm a}^{2}}{\bar{U}_{\rm t}' \bar{n}} ~ \left ( 1 - \frac{\bar{\gamma}' \bar{n}}{\bar{\gamma}} - \frac{\bar{U}_{\rm t}'' \bar{n}}{\bar{U}_{\rm t}'} + \frac{\bar{U}'' \bar{n}}{\bar{U}'} \right )~.
\end{equation}
An alternative route to the calculation of the photoresponse in the broadband limit is discussed in the rest of this section.

First, in the Euler equation~(\ref{eq:euler}) we neglect both terms $\gamma(x,t) \partial_{t} n(x,t)$ and $n(x,t) \partial_{t} \gamma(x,t)$ with respect to the product $\gamma(x,t) n(x,t) / \tau$.
This approximation is justified by substituting $\partial_{t}$ with $\omega$ and using the limit (i) above.
For consistency, in the same limit we also neglect the terms which are quadratic in the velocity.
Indeed, by dimensional considerations, we expect the amplitude of the velocity squared to be proportional to $|\omega / k(\omega)|^{2} \propto \omega \tau \ll 1$.
In Eq.~(\ref{eq:responsefunction}), the contribution from the terms quadratic in the velocity is multiplied by $\beta(\omega)$, which indeed vanishes in the limit (i).
With these approximations, the Euler equation yields
\begin{equation}
v(x,t) = -\frac{\tau e}{m \gamma(x,t)} \partial_{x} U(x,t)~.
\end{equation}
As discussed at the end of Sec.~\ref{sect:elstaticslinear}, we can rewrite this equation as
\begin{eqnarray}
v(x,t) & = & -\frac{\tau e}{m \gamma(x,t)} \left ( \frac{d U}{d n} \right )_{n=n(x,t)} \times \nonumber \\
& & \left ( \frac{d U_{\rm t}}{d n} \right )_{n=n(x,t)}^{-1} \times \partial_{x} U_{\rm t}(x,t)~.
\end{eqnarray}
Multiplying both sides of the equation by $-e n(x,t)$, introducing the charge current density $J(x,t) = -e n(x,t) v(x,t)$, and recognizing that $E(x,t) = \partial_{x} U_{\rm t}(x,t)$ is the electric field due to the external perturbation, we find the Ohm's law $J(x,t) = \sigma_{\rm t}(x,t) E(x,t)$ with the effective conductivity
\begin{equation}\label{eq:sigmat}
\sigma_{\rm t}(x,t) = \frac{\tau e^{2} n(x,t)}{m \gamma(x,t)} \left ( \frac{d U}{d n} \right )_{n=n(x,t)} \times \left ( \frac{d U_{\rm t}}{d n} \right )_{n=n(x,t)}^{-1}~.
\end{equation}
We point out that $\sigma_{\rm t}(x,t)$ is a function of electrostatic variables and thus, as discussed in Sec.~\ref{sect:elstaticslinear}, its space- and time-dependence can be rewritten as $\sigma_{\rm t}(x,t) = \sigma_{\rm t}(U_{\rm t}(x,t))$.

The second step of the derivation is to insert the expression for $v(x,t)$ into the continuity equation, where we also rewrite
\begin{eqnarray}
\partial_{t} n(x,t) & = & -e \left ( \frac{d n}{d U_{\rm t}} \right )_{U_{\rm t} = U_{\rm t}(x,t)} \times \partial_{t} U_{\rm t}(x,t) \nonumber \\
& \equiv & \tilde{C}_{\rm t}(U_{\rm t}(x,t))  \partial_{t} U_{\rm t}(x,t)~,
\end{eqnarray}
where $\tilde{C}_{\rm t}$ is an effective capacitance per unit area.
The continuity equation then reads
\begin{eqnarray}
& & - \tilde{C}_{\rm t} (U_{\rm t}(x,t)) \times \partial_{t} U_{\rm t}(x,t) + \sigma_{\rm t}(U_{\rm t}(x,t)) \times \partial_{x}^{2} U_{\rm t}(x,t) \nonumber \\
& & + \left ( \frac{d \sigma_{\rm t}}{d U_{\rm t}} \right )_{U_{\rm t} = U_{\rm t}(x,t)} \times \left ( \partial_{x} U_{\rm t}(x,t) \right )^{2} = 0~.
\end{eqnarray}
This is a diffusion equation for $U_{\rm t}(x,t)$, which has to be solved together with the BCs~(\ref{eq:bcsource}) and~(\ref{eq:bcdrain}).
This equation is identical to Eq.~(3) in Ref.~\cite{sakowicz_jap_2011} and its solution is discussed there.
The expression for the photovoltage, in the long-device-limit (ii), reads~\cite{sakowicz_jap_2011}
\begin{equation}
\Delta U_{\rm t} = \frac{1}{4} U_{\rm a}^{2} ~ \frac{1}{\sigma_{\rm t}(\bar{U}_{\rm t})} \left ( \frac{d \sigma_{\rm t}}{d U_{\rm t}} \right )_{U_{\rm t} = \bar{U}_{\rm t}}~.
\end{equation}
Deriving the expression~(\ref{eq:sigmat}) we find
\begin{equation}
\frac{1}{\sigma_{\rm t}(\bar{U}_{\rm t})} \left ( \frac{d \sigma_{\rm t}}{d U_{\rm t}} \right )_{U_{\rm t} = \bar{U}_{\rm t}} = \frac{1}{\bar{U}_{\rm t}' \bar{n}} ~ \left ( 1 - \frac{\bar{\gamma}' \bar{n}}{\bar{\gamma}} - \frac{\bar{U}_{\rm t}'' \bar{n}}{\bar{U}_{\rm t}'} + \frac{\bar{U}'' \bar{n}}{\bar{U}'} \right )
\end{equation}
and thus we exactly recover the result~(\ref{eq:dubroadband}).

\begin{acknowledgments}
Useful discussions with D.~Bandurin and A.~Principi are gratefully acknowledged.
This work was supported by the European Union's Horizon 2020 research and innovation programme under grant agreement no.~881603 - GrapheneCore3.
\end{acknowledgments}

\end{document}